\documentclass[twocolumn,11pt]{article}
\usepackage{vmargin}
\setmarginsrb{1.15in}{1.1in}{1.in}{2.0in}{0pt}{0pt}{0pt}{0pt}
\begin{document}

\title{Effectiveness of group interaction on conceptual standardized test performance}
\author{Chandralekha Singh, University of Pittsburgh}
\date{ 
We analyze the effectiveness of working in pairs on the Conceptual Survey of Electricity and Magnetism test in a
calculus-based introductory physics course. The group performance shows large normalized gain and evidence for
co-construction.
We discuss the effect of pairing students with different individual achievements.}

\maketitle

\vspace*{-.7in}
\section{Introduction}
\vspace{-.15in}

Peer collaboration as a learning tool has been exploited 
in many diverse instructional settings 
and with different types and levels of student populations. 
Here we investigate the effectiveness of working in pairs on the performance on a standardized conceptual multiple-choice test, 
Conceptual Survey of Electricity and Magnetism (CSEM), in a
calculus-based introductory physics course. 
In a double class period (1 hour 50 minutes), students were administered the 
test individually and in groups of two after instruction in the relevant concepts. 
In both instances (whether group or individual), students were allowed 50 minutes.
Between the individual and group testing there was a short 5-10 minute break, and students were 
required to turn in their first response so that they could not refer to it when working in group.
The test answers were never discussed with students so when they switched from individual to group (or vice versa), 
they did not know if their initial responses were correct.

Although some studies show that heterogenous groups are more effective for group learning, others show
that working with friends has special advantages. In this study, students were allowed to choose their own partners.
They were encouraged to discuss the response with each other, and each test counted for one quiz grade. 
Students had an additional incentive to discuss the concepts, because an examination in two weeks covered the same material. 
Moreover, students had extensive experience working with two peers in the recitation on context-rich 
problems and with one peer during the lecture on Mazur-style concept tests.
We note that the peer collaboration was unguided in that there was no help or facilitation from the instructor.

\vspace*{-.26in}
\section{Discussion}
\vspace*{-.16in}

To obtain two random equivalent samples,
all students in the class sitting on one side of the aisle took the individual test first, followed by the group test 
(IG treatment: 74 students or 37 pairs) while those on the other side of the aisle took the group test first before 
taking it individually (GI treatment: 54 students or 27 pairs). 
One reason for giving the test in both orders was to assess the effect of thinking individually before the peer discussion.
In the Mazur-style peer instruction, students are first asked to think about the concepts 
based on the assumption that not allowing an opportunity to think individually may prevent 
students from evaluating their own stand on a concept.
Another reason for designing both the IG and GI treatments was to evaluate the ``test-retest" or ``practice" effect.

Although the trends in some individual test items are interesting, here we only discuss
the effect of group work on the overall test scores.
In Table 1, we show the average individual and group scores for the GI and IG treatments and for the IG treatment, also the average normalized gain, 
$ g  =\langle (s_f-s_i)/(100-s_i) \rangle$, where $s_i$ and $s_f$ are the earlier (individual) and later (group) test scores in percent.
The fact that the group performance on the IG and GI treatments are virtually indistinguishable 
suggests that giving students an opportunity to think individually before the peer discussion did not improve their group performance.
Also, in the IG treatment, the normalized gain of $0.39$ in the group work is clearly not a ``test-retest" effect because
considering the treatment samples to be equivalent, we can compare the individual performance on IG treatment ($55\%$) with the 
group performance of GI treatment ($71.7\%$) which shows a gain of $0.37$ (indistinguishable from $0.39$).
Therefore, most of the following discussion will focus on the IG treatment. 

\begin{table}[h]
\centering
\begin{tabular}[t]{|c|c|c|c|}
\hline
 Treatment &I&G& $g$ \\[0.5 ex]
\hline \hline
GI&$70.3\%$& $71.7\%$&-\\[0.5 ex]
\hline
IG&$55\%$& $72.5\%$&$0.39$\\[0.5 ex]
\hline
\end{tabular}
\caption{{\it The average individual (I) and group (G) scores for the GI and IG treatments and $ g $ for
the IG treatment.}}
\label{junk9}
\end{table}

\vspace*{-.15 in}
But before moving on, we note some interesting trends in the time students took during the two successive testings in the IG and GI treatments. 
In the GI treatment, during the group work only, and in the IG treatment, both during the individual and group work, students 
took roughly the same amount of time.
One the other hand, students working individually after the group work in the GI treatment on an average took roughly one third
of the time spend on group work.
It appears that in the IG treatment, despite having worked on the problems individually, students
were willing to spend the time discussing the same test because they found peer collaboration useful.
However, in the GI treatment, after having discussed the test with peers, students were reasonably sure
about their thoughts and did not consider it necessary to brood over the problems again.

\vspace*{-.17in}
\subsection{Evidence for co-construction}
\vspace*{-.09in}

Although there is no consensus in the research literature on the definition of ``co-construction", we use the term here to 
denote cases where neither student alone chose the correct response but did as a group.
Co-construction can occur for several reasons. For example, if the group members chose {\it different} incorrect responses, they will have to
explain their reasoning to each other. This can unravel problems in their initial logic and complementary information provided by peers can 
help them converge on the correct solution. Even in cases where both students have the {\it same} incorrect response, co-construction can occur 
if students are unsure about their initial response and are willing to discuss their apprehensions with peers. 
Important clues provided by peers during the discussion can trigger recall of relevant concepts and can help the group co-construct.
One attractive feature of peer collaboration is that since both peers 
have recently gone through similar difficulties in assimilating and accommodating the new material,
they can often relate to each other's difficulties more easily than the instructor. 
The instructors' extensive experience can often make a concept so obvious and automatic that they may not 
comprehend why students are misinterpreting various aspect of a concept or finding them confusing.

Table 2 displays the percentage of the overall cases where neither, one, or both group members had the correct response individually
and how they changed during the group work (for IG treatment).  It shows the possibility for co-construction in $29\%$ of the cases where
neither group members were correct individually.

\begin{table}[h]
\begin{tabular}[t]{|c|c|c|c|}
\hline
\multicolumn{2}{|c|}{Individual Response } &\multicolumn{2}{|c|}{Group Response}{}\\[0.5 ex]\cline{3-4}
\multicolumn{2}{|c|}{} &Incorrect&Correct\\[0.5 ex]
\hline \hline
neither correct& $26\%$& $71\%$&$29\%$\\[0.5 ex]
\hline
one correct& $38\%$& $22\%$&$78\%$\\[0.5 ex]
\hline
both correct& $36\%$& $0\%$&$100\%$\\[0.5 ex]
\hline
\end{tabular}
\caption{{\it Distribution of the average group response for various combinations of individual response of group members in the IG treatment. 
The second column displays the percentage of the overall cases where neither, one, or both group members had the correct 
response individually.
}}
\end{table}

\vspace*{-.1 in}
To ensure that the cases in which both students individually chose the incorrect response but the group chose 
the correct response is not due to ``just guessing", we analyze the first row of Table 2 in detail. 
In Table 3, we subdivide this row based upon whether both partners had the same or different incorrect 
responses and if the group response was one of the original incorrect response or a third incorrect response.

\begin{table}[h]
\centering
\begin{tabular}[t]{|c|c|c|c|}
\hline
when both individual &\multicolumn{3}{|c|}{Group Response}{}\\[0.5 ex]\cline{2-4}
responses were incorrect&1&0 &$0^{\prime}$\\[0.5 ex]
\hline \hline
same incorrect($42\%$)&22$\%$&70$\%$&8$\%$\\[0.5 ex]
\hline
different incorrect($58\%$)&34$\%$&52$\%$&14$\%$\\[0.5 ex]
\hline
\end{tabular}
\caption{{\it Distribution of the average group response for cases where both members had the same or different incorrect 
individual response.
$1$, $0$ and $0^\prime$ refer to ``correct", ``one of the original incorrect" and ``a third incorrect" response respectively.
}}
\end{table}

Table 3 shows that in $22\%$ of the cases when both group members had the same incorrect response, and in $34\%$ of the cases
when both had different incorrect response, the group response was correct.
In comparison, a relatively small frequency of a third incorrect group response that was not originally selected by 
either members suggests that students were not ``just guessing" (see Table 3).
Although we did not conduct formal interviews with students after they worked in groups, 
we briefly discussed the aspects of the group work they found helpful with several students. Most students admitted that they got
useful insights about various electricity and magnetism concepts by discussing them with peers. Students frequently noted (often with
examples) that they had difficulty interpreting the problems alone but interpretation became easier with a friend. 
They also said that talking to peers forced them to think harder about the concepts, find fault with their initial reasoning, and
reminded them of concepts they had difficulty recalling on their own.
Qualitative observations show that students were more likely to draw field lines, write equations or scribble on their exams in 
the group work than in the individual work.

\vspace*{-.18in}
\subsection{Negative impact of grouping? }
\vspace{-.09in}

Table 2 (second row) shows that out of all the cases in which one group member individually had the correct response and the
other had an incorrect response, $78\%$ of the group responses were correct. The fact that $22\%$ of such cases
resulted in an incorrect group response is not very troublesome because it was an unguided peer
discussion. It can happen if students who individually chose the correct response are not very confident and cannot defend
or justify their response. The fact that a majority of such cases resulted in correct group responses suggests that students
who individually chose the correct response were generally more confident and were able to justify their
choice. It also suggests that students who picked an incorrect item individually were less sure of their choice and
were willing to listen to their peers. As will be discussed in the next section, group work always results in 
individual gain.

\vspace*{-.2in}
\subsection{Individual gain and retention}
\vspace{-.1in}

The {\it individual} performance in the GI treatment was much superior ($70.3\%$) compared to the IG treatment ($55\%$).
One can hypothesize that students could immediately recall the group responses for all the 32 test items in the GI treatment and their superior 
performance does not reflect the effectiveness of group work with regard to the retention of the concepts discussed. 
Similarly, in the IG treatment, the superior group performance compared to the individual performance is due to a large number of cases 
where the group member with the
correct response was able to convince the one with the incorrect response. It does not guarantee that students 
will retain what they learned in the group work. To investigate the impact of group interaction on retention, two weeks 
after the IG and GI treatments, all students individually took the CSEM test again. 
In view of the fact that students earlier went through either the IG or GI treatments, we now relabel the treatments IGI and GII.
We note that 
students did not know that they will be taking the same test again.
Although it would have been better to use different, equally reliable test for assessing similar concepts; unfortunately, none was available.

In the IGI treatment, the average score for the second individual testing was $74\%$, a gain of
$0.42$ compared to the initial individual score of $55\%$. 
A detailed comparison of the group vs. the second individual test scores shows that $81\%$ of
the overall individual responses chosen by the members of a particular group were the same as the group responses. Out of the $19\%$ 
individual responses that were different from the group, roughly $11\%$ went from incorrect to correct while $8\%$ went from
correct to incorrect. 
There are two competing effects: the fact that students forgot some group responses over time and the fact 
that they had time to study for the second individual test; a large fraction of group response was retained even after two weeks.
The corresponding numbers in the GII treatment are virtually the same. 

\vspace{-.20in}
\subsection{Effective pairing}
\vspace{-.10in}

To learn about the type of pairings that will optimize the overall gain for this conceptual test,
we divided the 74 students in the IGI treatment into three equal types: high (A), middle (B), and low (C), based upon their initial individual
score on CSEM. In Tables 4a and 4b, we show the average initial individual score (left) and the normalized gain in the second 
individual testing (right) for all the nine possible pairs. 
In comparison, for all 74 students together, the average initial individual score was $55\%$ and the average {\it individual} gain was $0.42$.

\begin{table}
\mbox{
\hspace*{-0.13in}
\begin{minipage}{0.3 \linewidth}
\begin{tabular}[t]{|c|c|c|c|}
\hline
(a)&\multicolumn{3}{|c|}{pairing}\\[0.5 ex]\cline{2-4}
&A &B &C \\[0.5 ex]
\hline \hline
A&73&74&71\\[0.5 ex]
\hline
B&56&54&56\\[0.5 ex]
\hline
C&34&36&36\\[0.5 ex]
\hline
\end{tabular}
\end{minipage}
}
\hspace*{0.36in}
\begin{minipage}{0.3 \linewidth}
\begin{tabular}[t]{|c|c|c|c|}
\hline
(b)&\multicolumn{3}{|c|}{pairing}\\[0.5 ex]\cline{2-4}
&A &B &C \\[0.5 ex]
\hline \hline
A&0.54&0.55&0.29\\[0.5 ex]
\hline
B& 0.65&0.51&0.26\\[0.5 ex]
\hline
C&0.41&0.40&0.37\\[0.5 ex]
\hline
\end{tabular}
\end{minipage}
\caption{{\it (a) [left] The average initial individual score, and (b) [right] the normalized gain for the nine types of
pairs. The top row of the Table refers to the performance of A students for different kinds of pairings: (A A),  (A B) and (A C).
Similarly, the middle and bottom row refer to the performance of B and C students respectively. 
}}
\end{table}

Table 4 shows that although all students in the IGI treatment gained in the second individual testing two weeks later compared to the 
first, the gain matrix is not symmetric. 
For example, the gain of A student due to the interaction with C ($0.29$) is not the same as that of C due to A 
($0.41$) in the (A C) pairing. It is striking that the gain of A and B students is significantly lower when they paired
with C students. `A' students have similar gains whether they paired with A or B students while B students
benefit more from pairing with A than with another B student.
Interestingly, the gain of C students (lowest third) is virtually the same regardless of who they paired with 
(bottom row of Table 4). One hypothesis is that while pairing with a higher achievement student helped C students do well
in the group test, they did not retain all of the concepts because they were subdued by the higher achieving student and did not participate 
actively in the discussions. 
Thus, the opportunity to learn from the higher achieving student may have been outweighed by the inability of C students to 
participate fully in the discussions and process the information at the rate discussed by the other student.
On the other hand, in the (C C) pairing, both students had comparable but at least some complementary 
knowledge and both actively participated in the discussions.  
The evidence for this hypothesis comes from the comparison of the average group (left) and second individual test scores (right) for
each of the pairings as shown in Tables 5a and 5b. 

\begin{table}
\mbox{
\hspace*{-0.07in}
\begin{minipage}{0.3 \linewidth}
\begin{tabular}[t]{|c|c|c|c|}
\hline
(a)&\multicolumn{3}{|c|}{pairing}\\[0.5 ex]\cline{2-4}
&A &B &C \\[0.5 ex]
\hline \hline
A& 88&85&72\\[0.5 ex]
\hline
B &85&73&62\\[0.5 ex]
\hline
C&72&62&54\\[0.5 ex]
\hline
\end{tabular}
\end{minipage}
}
\hspace*{0.5in}
\begin{minipage}{0.3 \linewidth}
\begin{tabular}[t]{|c|c|c|c|}
\hline
(b)&\multicolumn{3}{|c|}{pairing}\\[0.5 ex]\cline{2-4}
&A &B &C \\[0.5 ex]
\hline \hline
A&88&88&79\\[0.5 ex]
\hline
B&85&78&67\\[0.5 ex]
\hline
C&61&62&60\\[0.5 ex]
\hline
\end{tabular}
\end{minipage}
\caption{{\it (a) [left] The average group score, and (b) [right] the second individual test score for the nine types of pairs. 
}}
\end{table}

A comparison of Tables 5a and 5b shows that the average individual score of C students in the (A C) pairing
after two weeks ($61\%$) is lower compared to their group score ($72\%$).
Also, a comparison of Tables 4a and 5a shows that A students did not benefit from interactions
with C students and the average initial individual score for A students and the group score for the (A C) 
pairing are the same. Similar comparisons for B students shows that although they benefitted from all 
types of interactions, their gain improved as they interacted with higher achieving students.
It appears that at least for this conceptual test, the pairing that helps maximize the
overall gain is one that only has (A B) and (C C) pairs. It will
be useful to investigate the extent to which this result is universal, i.e., whether two peers collaborating 
on conceptual tests show highest overall gains
when the high and middle achievement students are paired, and
the low achievement students are paired with each other.

\end{document}